\documentclass[%
superscriptaddress,
twocolumn,
showpacs,
amsmath,amssymb,
aps,
prl,
floatfix,
showkeys,
reprint,
]{revtex4-1}

\usepackage{graphicx}
\usepackage{dcolumn}
\usepackage{bm}
\usepackage{amsmath}
\usepackage{amssymb}
\usepackage{multirow}
\usepackage{color}
\usepackage{booktabs}
\usepackage{tabularx}
\usepackage[english]{babel}
\usepackage[utf8]{inputenc}
\usepackage{enumerate}
\usepackage[cmyk,dvipsnames]{xcolor}
\definecolor{pacificb}{HTML}{1CA9C9}

\usepackage[normalem]{ulem}

\begin{document}
	\title{Role of higher-order exchange interactions for skyrmion stability}
	
	\author{Souvik Paul}
	\email{paul@physik.uni-kiel.de}
	\affiliation{Institute of Theoretical Physics and Astrophysics, University of Kiel, Leibnizstrasse 15, 24098 Kiel, Germany}
	
	\author{Soumyajyoti Haldar}
	\affiliation{Institute of Theoretical Physics and Astrophysics, University of Kiel, Leibnizstrasse 15, 24098 Kiel, Germany}
	
	\author{Stephan von Malottki}
	\affiliation{Institute of Theoretical Physics and Astrophysics, University of Kiel, Leibnizstrasse 15, 24098 Kiel, Germany}
	
	\author{Stefan Heinze}
	\affiliation{Institute of Theoretical Physics and Astrophysics, University of Kiel, Leibnizstrasse 15, 24098 Kiel, Germany}
	
	\date{\today}

\begin{abstract}
Transition-metal interfaces and multilayers are a promising class of systems to realize nanometer-sized, stable magnetic skyrmions for future spintronic devices. For room temperature applications, it is crucial to understand the interactions which control the stability of isolated skyrmions. Typically, skyrmion properties are explained by the interplay of pair-wise exchange interactions, the Dzyaloshinskii-Moriya interaction and the magnetocrystalline anisotropy energy. Here, we demonstrate that the higher-order exchange interactions -- which have so far been neglected -- can play a key role for the stability of skyrmions. We use an atomistic spin model parametrized from first-principles and compare three different ultrathin film systems. We consider all fourth order exchange interactions and show that, in particular, the four-site four spin interaction has a giant effect on the energy barrier preventing skyrmion and antiskyrmion collapse into the ferromagnetic state. Our work opens new perspectives to enhance the stability of topological spin structures.

\end{abstract}

\maketitle
Magnetic skyrmions -- localized spin structures with a topological charge \cite{tokura13} -- have raised high hopes for future magnetic memory and logic devices due to their nanoscale dimensions, stability and ultra-low energy driven motion \cite{tomasello14,zhou14,junichi13,sampaio13,fert13}. Skyrmion lattices have been first observed in bulk magnets with a broken inversion symmetry in their crystal structure \cite{Muehlbauer2009,Yu2010a}. The discovery of a skyrmion lattice in a single atomic layer of Fe on the Ir(111) surface \cite{Heinze2011} has opened the door to a new class of systems: transition-metal interfaces and multilayers. Due to the possibility of varying film composition and structure, these systems allow to modify magnetic interactions and thereby the properties of skyrmions. At such transition-metal interfaces, individual magnetic skyrmions with diameters ranging from a few 100 nanometers down to a few nanometers have been realized as a metastable state in the field-polarized ferromagnetic background \cite{Yu2010a,Romming2013,Romming2015,Moreau-Luchaire2016,Soumyanarayanan2017,hsu19,wilson14,meyer19} as needed for applications.

A key challenge of skyrmion based data processing and storage technology is the robustness of information carriers, i.e., stability of the skyrmionic bits, against random thermal fluctuations at operating temperatures. At finite temperature, the magnetic moments of skyrmions are coupled to the environment, which induces fluctuations. Over time, a rare energy fluctuation can grow in excess of the barrier height and can prompt the skyrmion to overcome the barrier and collapse to the ferromagnetic background leading to a loss of topological charge. Therefore, an accurate assessment of barrier height is essential to determine the stability of skyrmions. To achieve data reading and writing capabilities of skyrmionic bits with high efficiency, a control over the barrier height is also necessary. 

The existence of chiral magnetic skyrmions \cite{Bogdanov1989} is ascribed to a competition of the Heisenberg pair-wise exchange interaction, the Dzyaloshinskii-Moriya interaction (DMI) \cite{Dzyaloshinskii1957,Moriya1960}, the magnetocrystalline anisotropy and the dipole-dipole interactions. A prerequisite of DMI -- which provides a unique rotational sense to skyrmions -- is the concerted action of spin-orbit coupling (SOC) and broken inversion symmetry, which can be achieved at interfaces of transition-metals \cite{Bode2007}. The DMI further stabilizes metastable isolated skyrmions against annihilation into the ferromagnetic background \cite{Bogdanov1989,Bogdanov1994}. Often the exchange interactions are treated in a micromagnetic or effective nearest-neighbor approximation. However, the exchange interactions are long-range in itinerant magnets such as $3d$ transition-metals. This can lead to a competition between exchange interactions from different shells of atoms resulting in an enhanced skyrmion stability \cite{Malottki2017a,meyer19} even in the absence of DMI \cite{Leonov2015}.

The itinerant character of $3d$ transition-metals limits the applicability of the Heisenberg model to describe their magnetic properties. Based on the spin-$1/2$ Hubbard model, it has been shown that the higher-order exchange interactions (HOI) beyond pair-wise Heisenberg exchange can arise such as the two-site four spin (biquadratic) or the four-site four spin interaction \cite{takahashi77,MacD88}. Such higher-order terms can lead to intriguing magnetic ground states due to a superposition of spin spirals -- so-called multi-$Q$ states -- which have been predicted based on first-principles calculations \cite{Kurz2001}. The interplay of the four-site four spin interaction and DMI is the origin of the nanoskyrmion lattice of the Fe monolayer on Ir(111) \cite{Heinze2011}
and the effect of the biquadratic interaction on skyrmion lattice formation has been studied systematically \cite{Hayami2017}.
It has been further demonstrated that the HOI can compete with the DMI and stabilize novel magnetic ground states \cite{romming18}. Based on a multi-band Hubbard model, a three-site four spin interaction has recently been proposed for systems with a spin beyond $S$= $1/2$ in addition to the biquadratic and the four-site four spin interaction in fourth order perturbation theory of the hopping parameter $t$ with respect to the Coulomb energy $U$ \cite{hoffmann18}. This term has been attributed to 
stabilize a double-$Q$ or so-called up-up-down-down ($uudd$) state in an Fe monolayer on Rh(111) \cite{Kronelein2018}. Despite the compelling experimental evidence of the relevance of HOI, they have been neglected so far in the theoretical description of the properties of isolated 
magnetic skyrmions at transition-metal interfaces.

\begin{figure*}[htbp!]
	\centering
	\includegraphics[scale=1.0]{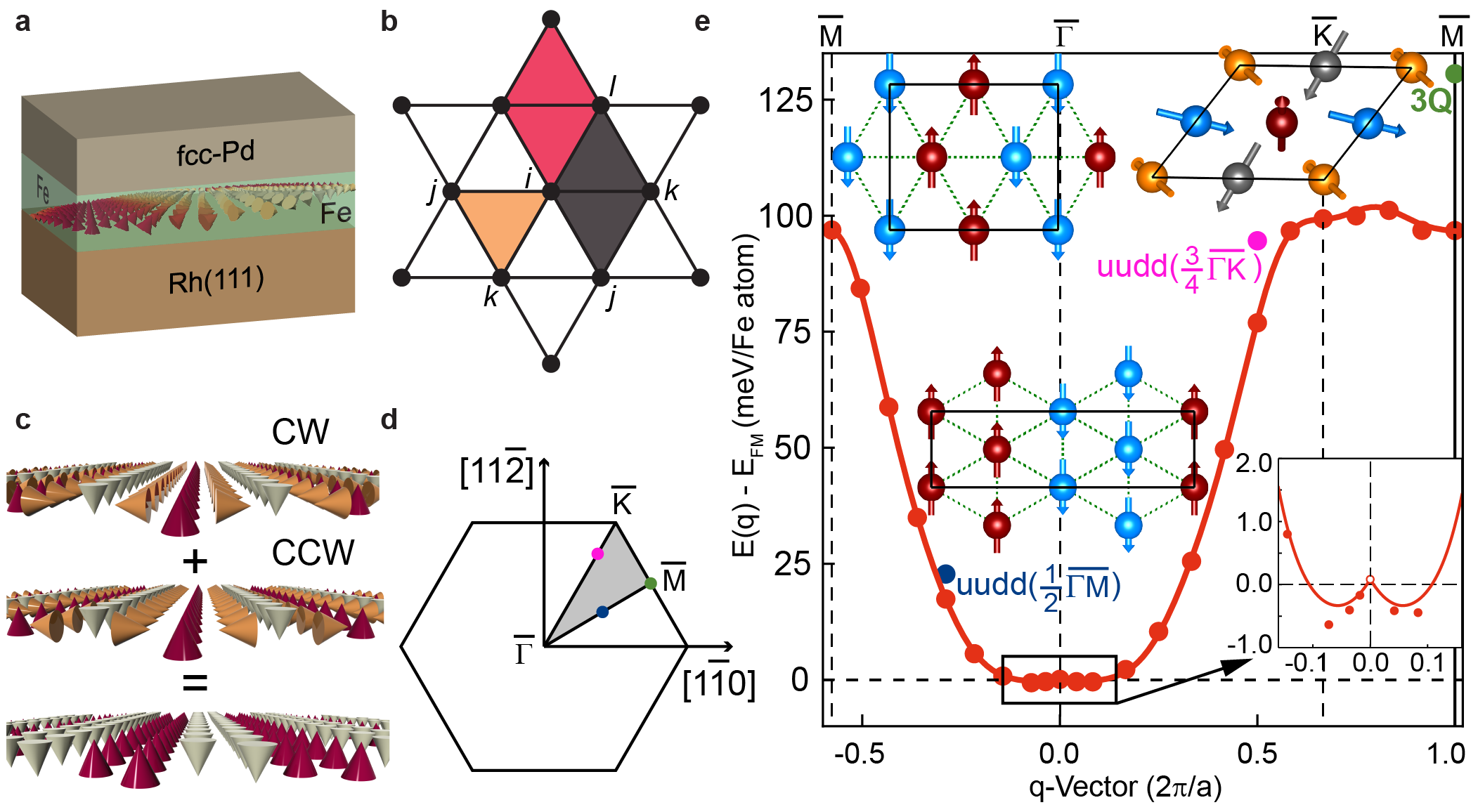}
	\caption{\textbf{Higher-order exchange interactions and multi-$Q$ states}. \textbf{a} Illustration of the Pd/Fe/Rh(111) ultrathin film with a spin spiral propagating in the Fe layer. \textbf{b} Four sites ($ijkl$) involved in the four-site four spin interaction form a diamond (red and brown) on a 2D hexagonal lattice and three sites ($ijk$) involved in the three-site four spin interaction result in a triangle (yellow). A total of 12 diamonds and 6 triangles are possible for site $i$. The 12 diamonds can be categorized into two groups. One from each group is shown in red and brown. All the three sites $j$, $k$ and $l$ of the brown diamond are nearest neighbors to the central site $i$. The topmost site of the red diamond is a next-nearest neighbor to $i$, while the other two remain nearest neighbors. \textbf{c} Formation of an $uudd$ state as a superposition of two 90$^\circ$ spin spirals with opposite rotational sense, i.e., clockwise (CW) and counterclockwise (CCW). \textbf{d} 2D Brillouin zone with two high symmetry direction $\overline{\Gamma \mathrm{K}}$ and $\overline{\Gamma \mathrm{M}}$. The $\mathbf{q}$ vectors corresponding to the $uudd$ state along $\overline{\Gamma \mathrm{K}}$ (pink filled circle) and along $\overline{\Gamma \mathrm{\mathrm{M}}}$ (blue filled circle) as well as the $3Q$ state at the $\overline{M}$ point (green filled circle) are indicated. \textbf{e} Energy dispersion $E(\mathbf{q})$ of homogeneous spin spirals of Pd/Fe/Rh(111). The filled circles (red) are DFT total energies including spin-orbit coupling. The solid line is a fit to the Heisenberg and the Dzyaloshinskii-Moriya interaction. The energies of the $uudd$ state along $\overline{\Gamma \mathrm{K}}$ (pink filled circle), the $uudd$ state along the $\overline{\Gamma \mathrm{M}}$ direction (blue filled circle) and the 3$Q$ state (green filled circle) are denoted at the $\mathbf{q}$ value of the corresponding single-$Q$ state. The spin structures of the $uudd$ state along $\overline{\Gamma \mathrm{K}}$ (pink), which is formed by a superposition of two 90$^{\circ}$ spin spirals at $\textbf{q}$=$\pm (3/4)\overline{\Gamma \mathrm{K}}$, the $uudd$ state along the $\overline{\Gamma \mathrm{M}}$ direction (blue), which is formed by a superposition of two 90$^{\circ}$ spin spirals at $\textbf{q}$=$\pm (1/2)\overline{\Gamma \mathrm{M}}$, and the 3$Q$ state (green), which is a superposition of three spin spirals corresponding to three $\overline{\mathrm{M}}$ points (green) in 2DBZ. Inset of \textbf{e} shows that the ground state of Pd/Fe/Rh(111) is a spin spiral with a wavelength of 4.8 nm.}
	\label{fig:fig1}
\end{figure*}

Here, we reveal the intriguing role played by the HOI for the stability of topologically non-trivial spin structures such as skyrmions and antiskyrmions at transition-metal interfaces. We use spin dynamics simulations based on an atomistic spin model with all parameters calculated via density functional theory (DFT). The energy barrier, preventing skyrmions and antiskyrmions to collapse into the ferromagnetic state, is obtained using the geodesic nudged elastic band (GNEB) method \cite{Bessarab2015}. We consider three ultrathin film systems: (i) fcc-Pd/Fe bilayer on Rh(111) (Fig.~\ref{fig:fig1}a) for which sub-10 nm skyrmions have been predicted at low magnetic field \cite{som18}, (ii) fcc-Pd/Fe bilayer on Ir(111), the most intensively studied ultrathin film system that hosts isolated skyrmions \cite{Romming2013,Romming2015,Hagemeister:15.1,Kubetzka:17.1,Hanneken:15.1,Bessarab2018,Malottki2017a,Dupe2014,boettcher18,levente16.93,levente17.95} and an hcp-Fe/Rh bilayer on Re(0001) 
with an in-plane easy magnetization axis \cite{mine}.

Upon including the HOI, the stability of skyrmions and antiskyrmions in all of these films is greatly modified. Surprisingly, the effect of the biquadratic and the three-site four spin interaction concerning the energy barrier is to a good approximation already captured in the exchange constants 
obtained by mapping the DFT results to a spin model neglecting HOI.
The four-site four spin interaction has a large effect on the saddle point and it is responsible for the large change in energy barriers. We find a linear scaling of the barrier height with the four-site four spin interaction. The barrier is enhanced or reduced depending on its sign. Even small values of the four-spin four site interactions of 1 to 2 meV, typical for $3d$ transition metals, modify the energy barrier by 50 to 120 meV. This leads to a huge enhancement or reduction of the skyrmion or antiskyrmion lifetime. We further show that the HOI can stabilize topological spin structures in the absence of DMI.       

\section*{}
\noindent{\large{\textbf{Results}}}\par
\noindent{\textbf{Atomistic spin model and DFT calculations.}} We describe the magnetic state of an ultrathin film by a set of classical magnetic moments $\{\textbf{M}_{i}\}$ localized on each atom site $i$ of a hexagonal lattice and their dynamics is governed by the following Hamiltonian:

\begin{widetext}
\begin{eqnarray} \label{eq:hamiltonian}
\begin{aligned}
\mathcal{H} &=- \sum_{ij} J_{ij} (\textbf{m}_{i} \cdot \textbf{m}_{j}) - \sum_{ij} \textbf{D}_{ij} \cdot (\textbf{m}_{i}\times\textbf{m}_{j}) - \sum_{i} K(m^{z}_{i})^2 - \sum_{i} \mu_{s} \textbf{B} \cdot \textbf{m}_{i} - \sum_{ij} B_{ij} (\textbf{m}_{i} \cdot \textbf{m}_{j})^2 \\ &- 2 \sum_{ijk} Y_{ijk} (\textbf{m}_{i} \cdot \textbf{m}_{j}) (\textbf{m}_{j} \cdot \textbf{m}_{k}) - \sum_{ijkl} K_{ijkl} [(\textbf{m}_{i} \cdot \textbf{m}_{j}) (\textbf{m}_{k} \cdot \textbf{m}_{l}) +(\textbf{m}_{i} \cdot \textbf{m}_{l}) (\textbf{m}_{j} \cdot \textbf{m}_{k})-(\textbf{m}_{i} \cdot \textbf{m}_{k}) (\textbf{m}_{j} \cdot \textbf{m}_{l})]
\end{aligned}
\end{eqnarray}
\end{widetext}

where $\textbf{m}_{i}$=$\textbf{M}_{i}/M_{i}$ is a unit vector. The exchange constants ($J_{ij}$), the DMI vectors ($\textbf{D}_{ij}$), the magnetic moments ($\mu_{s}$) and the uniaxial magnetocrystalline anisotropy energy (MAE) constant ($K$) were calculated based on DFT (see Refs. \cite{Dupe2014,Malottki2017a,som18,mine}). We neglect the dipole-dipole interaction since it is small in ultrathin films, which is of the order of 0.1~meV/atom, and it can be effectively included into the MAE \cite{Draaisma:88.1,Lobanov:16.1}.

The last three terms are the biquadratic interaction ($B_{ij}$), the three-site four spin interaction ($Y_{ijk}$) and the four-site four spin interaction ($K_{ijkl}$), respectively. Since these terms arise from the fourth-order perturbation theory, we restrict ourselves to the nearest-neighbor approximation, i.e., up to the first term of these HOI. The corresponding constants are denoted as $B_{1}$, $Y_{1}$ and $K_{1}$ \footnote{Note the subscript of the biquadratic ($B_{1}$) and the four-site four spin ($K_{1}$) constants to distinguish them from the MAE ($K$) and the external magnetic field ($B$)}.
 
The evaluation scheme of the HOI on a hexagonal two-dimensional (2D) lattice is illustrated in Fig.~\ref{fig:fig1}b. Within the nearest-neighbor approximation, the minimal connection among the three ($ijk$) and four ($ijkl$) adjacent lattice sites on a hexagonal lattice leads to a triangle and a diamond, respectively. A clusters of 12 diamonds and 6 triangles for each lattice site contributes to the four-site and three-site four spin interactions, respectively. Out of the 12 diamonds, two distinct types, each containing six diamonds, can be identified (Fig.~\ref{fig:fig1}b). For the biquadratic interaction, 6 nearest-neighbor pairs, analogous to the nearest-neighbor exchange constant ($J_{1}$), are sufficient.

In order to obtain the exchange constants, the energy dispersion of homogeneous flat spin spirals is calculated using DFT. 
A spin spiral is characterized by a wave vector $\textbf{q}$ in the two-dimensional Brillouin zone (Fig. \ref{fig:fig1}d) and the magnetic moments of an atom at lattice site $\textbf{R}_{i}$ is given by $\textbf{M}_{i}$= $M$(sin($\textbf{qR}_{i}$),cos($\textbf{qR}_{i}$),0) with the size of the magnetic moment $M$. Figure \ref{fig:fig1}e shows the energy dispersion $E(\textbf{q})$ of  spin spirals along two high symmetry directions obtained via DFT for a fcc-Pd/Fe bilayer on the Rh(111) surface \cite{som18}. At the high symmetry points of the two-dimensional Brillouin zone (2DBZ), we find well-known magnetic states: a ferromagnetic (FM) state at the $\overline{\Gamma}$ point, a row-wise antiferromagnetic (AFM) state at the $\overline{\mathrm{M}}$ point and a N\'eel state with angles of 120$^\circ$ between adjacent magnetic moments at the $\overline{\mathrm{K}}$ point. 

Clearly, the FM state is the energetically lowest among these three states (Fig.~\ref{fig:fig1}e). Along both the symmetry directions, the 90$^\circ$ spin spirals (Fig.~\ref{fig:fig1}c) are found at $\textbf{q}=(1/2) \overline{\Gamma \mathrm{M}}$ and at $\textbf{q}=(3/4) \overline{\Gamma \mathrm{K}}$ (Fig.~\ref{fig:fig1}d). The total energy of homogeneous spin spirals without SOC is fitted to functions obtained by expressing the Heisenberg model in reciprocal space. When SOC is included, the DMI lowers the energy of cylcoidal spin spirals with a clockwise rotational sense in the vicinity of the $\overline{\Gamma}$ point  (see inset of Fig.~\ref{fig:fig1}e). The MAE, on the other hand, shifts the energy of spin spirals by $K$/2 with respect to the FM state.
 
Since the functional form of the three-site four spin and the biquadratic interactions for homogeneous spin spirals resemble that of the first three exchange constants, we cannot separate the exchange and the higher-order constants by fits (see Methods). Therefore, we calculate the higher-order exchange constants from the energy difference between the spin spiral (single-$Q$) and multi-$Q$ states and modify the exchange constants obtained from fit by neglecting the HOI accordingly.

The multi-$Q$ state is a superposition states of spin spirals corresponding to the symmetry equivalent $\mathbf{q}$ vectors (cf.~Fig.~\ref{fig:fig1}(c,d)). Within the Heisenberg model of pair-wise interaction, the multi-$Q$ and single-$Q$ states are energetically degenerate. However, in the extended-Heisenberg model (Eq.~(\ref{eq:hamiltonian})), the HOI lift the degeneracy which provides a way to compute their strengths. In DFT calculations, all the interactions are implicitly included through the exchange-correlation functional. Therefore, we can obtain the HOI constants from total energy calculations of multi-$Q$ and single-$Q$ states.

We consider two collinear states, the so-called $uudd$ or double-row wise antiferromagnetic states \cite{Hardrat2009} and a three-dimensional non-collinear state, the so-called 3$Q$ state \cite{Kurz2001}, to uniquely determine 
three higher-order exchange constants (for spin structures see insets of Fig.~\ref{fig:fig1}e). The resulting $uudd$ state along $\overline{\Gamma \mathrm{K}}$, represents rows of alternating spins in the nearest-neighbor direction, is formed by superposition of $\textbf{q}$=$\pm (3/4)\overline{\Gamma \mathrm{K}}$, whereas the other along $\overline{\Gamma \mathrm{M}}$, represents rows of alternating spins in the next nearest-neighbor direction, is formed by superposition of $\textbf{q}$=$\pm (1/2)\overline{\Gamma \mathrm{M}}$ (Fig.~\ref{fig:fig1}c). The 3$Q$ state -- a three-dimensional spin structure on the two-dimensional hexagonal lattice (inset of Fig.~\ref{fig:fig1}e) -- is constructed by a superposition of three spin spirals at the three symmetry equivalent $\overline{\mathrm{M}}$ points of the 2DBZ \cite{Kurz2001} (Fig.~\ref{fig:fig1}d).

The biquadratic ($B_{1}$), the three-site four spin ($Y_{1}$) and the four-site four spin interaction ($K_{1}$) constants are computed from the energy differences between the multi-$Q$ and the corresponding single-$Q$ states by solving the following equations \cite{hoffmann18}:
\begin{subequations} \label{eq:mqsq}
\begin{align}
E_{\overline{\mathrm{M}}}^{3Q} - E_{\overline{\mathrm{M}}}^{\mathrm{1Q}} = \frac{16}{3} (2K_{1}+B_{1}-Y_{1}) \label{eq6} \\
E_{\overline{\mathrm{M}}/2}^{uudd} - E_{\overline{\mathrm{M}}/2}^{1Q} = 4 (2K_{1}-B_{1}-Y_{1}) \label{eq7} \\
E_{3{\overline{\mathrm{K}}}/4}^{uudd} - E_{3{\overline{\mathrm{K}}}/4}^{1Q} = 4 (2K_{1}-B_{1}+Y_{1}) \label{eq8}
\end{align}
\end{subequations}

The three multi-$Q$ states are higher in energy compared to the corresponding spin spirals states for Pd/Fe/Rh(111) (Fig.~\ref{fig:fig1}e) and they are energetically far above the spin spiral ground state (energies are given in Supplementary Table 1). Nevertheless, they have a large effect on skyrmions in this film system as we show below. 
 
The computed higher-order constants modify the first three exchange constants, obtained 
from fits of the spin spiral energy dispersion neglecting HOI, 
as follows (see Methods for a detailed derivation):
\begin{subequations} \label{eq:impJs}
\begin{align}	
J^{\prime}_{1}&= J_{1}-Y_{1} \\
J^{\prime}_{2}&= J_{2}-Y_{1} \\
J^{\prime}_{3}&= J_{3}-B_{1}/2 
\end{align}
\end{subequations}
where we denote the exchange parameters obtained from fits neglecting HOI as unprimed and the modified ones by considering the higher-order terms as primed. It is important to note that the 
four-site four spin interaction does not adjust any exchange parameter since its 
contribution to the energy dispersion of spin spirals is a constant value of $-12 K_1$ independent of the spin spiral vector.

\begin{table} [!htbp]
	\centering
	\begin{ruledtabular}
		\begin{tabular}{ccccccc}
			$\mathrm{Systems}$ & $J^{\prime}_{1}$ & $J^{\prime}_{2}$ & $J^{\prime}_{3}$ & $B_{1}$ & $Y_{1}$ & $K_{1}$\\
			\colrule
			Pd/Fe/Rh(111) & 11.73 & $-4.31$ & $-$4.21 & 2.74 & 1.61 & 2.61 \\
			Pd/Fe/Ir(111) & 13.60 & $-3.28$ & $-4.17$ & 2.96 & 0.80 & 2.14 \\
			Fe/Rh/Re(0001) & 8.85 & $-0.77$ & 0.05 & $-0.39$ & 1.00 & $-1.36$ \\
		\end{tabular}
		\caption{\textbf{Exchange constants including HOI.} Exchange constants for $i$-th neighbor spins ($J^{\prime}_{i}$), biquadratic ($B_{1}$), three-site four spin ($Y_{1}$) and four-site four spin interaction $(K_{1}$) constants for Pd/Fe/Rh(111), Pd/Fe/Ir(111) and Fe/Rh/Re(0001). The exchange constants are modified upon including the HOI according to Eqs.~(\ref{eq:impJs}a-c). This table shows only those exchange constants which are modified upon including the HOI. The full sets of exchange constants as well as the DMI, the MAE and the total magnetic moments used in the atomistic spin dynamics simulations are listed in Supplementary Table 2 and Supplementary Table 3. All values are given in meV.}
		\label{tab:table1}
	\end{ruledtabular}
\end{table}

The exchange and the higher-order exchange constants of Pd/Fe/Rh(111) are displayed in Table~\ref{tab:table1}. We have
performed similar DFT calculations in order to obtain these constants for the other two ultrathin film systems: Pd/Fe/Ir(111) and Fe/Rh/Re(0001) (Table~\ref{tab:table1} and Supplementary Table 1). We find that the Pd/Fe bilayer on Rh(111) and on Ir(111) behaves similar in terms of exchange and higher-order exchange constants which can be understood
based on the fact that Rh and Ir are isoelectronic $4d$ and $5d$ transition metals. In these two film systems, the signs of the nearest-neighbor exchange constant ($J_1$) and the second and third nearest neighbors are opposite which leads to exchange frustration \cite{Malottki2017a,som18}. The exchange interaction in Fe/Rh/Re(0001), in contrast, is dominated by the nearest-neighbor exchange constant. Note that the sign of the biquadratic ($B_{1}$) and the four-site four spin constants ($K_{1}$) is negative in Fe/Rh/Re(0001), while it is positive for the other two systems. As we will see below, the sign of the four-site four spin interaction is essential for the skyrmion stability in these films.

\begin{figure*}[htbp!]
	\centering
	\includegraphics[scale=1.0,clip]{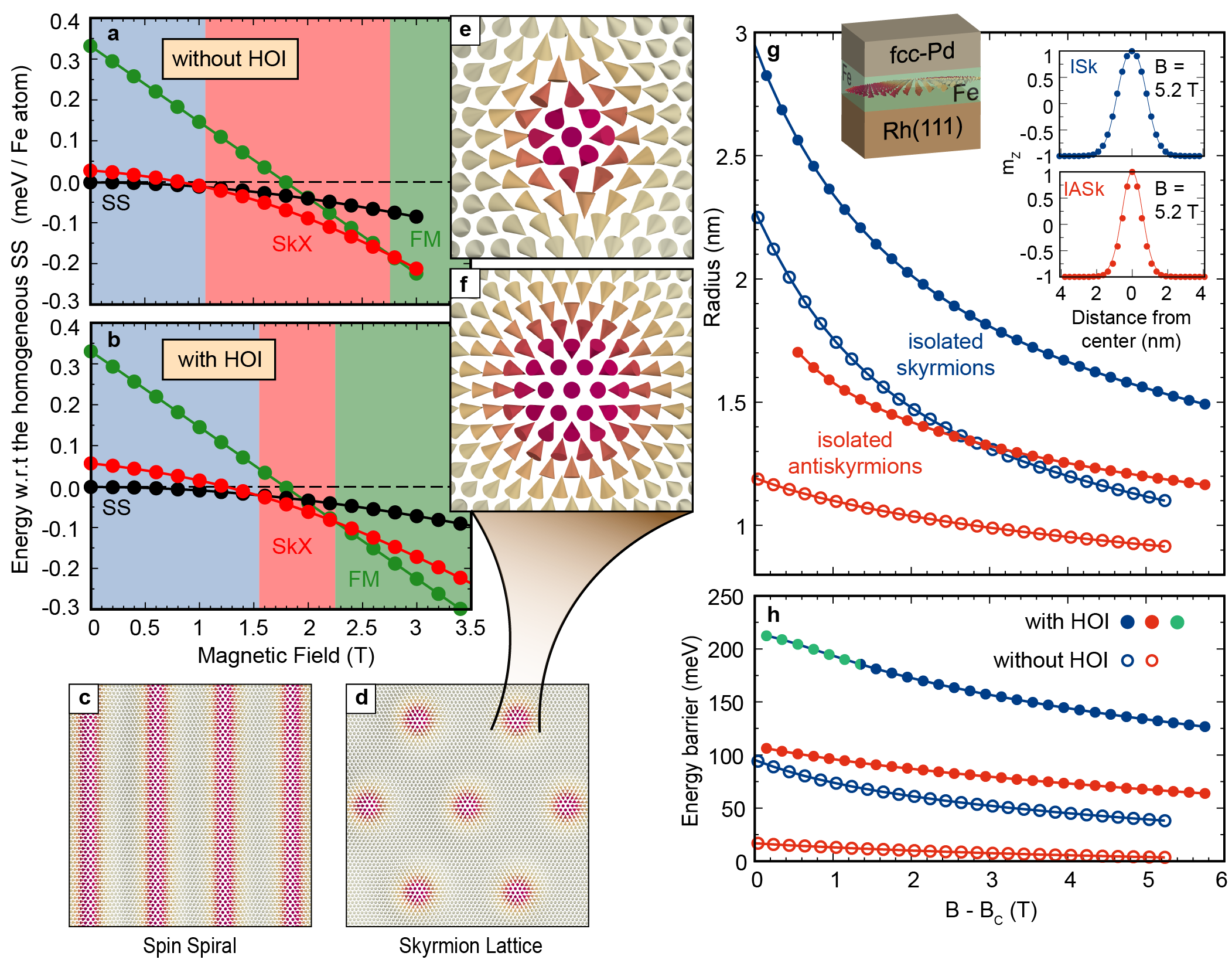}
	\caption{\textbf{Phase diagram, radius and barrier heights of Pd/Fe/Rh(111) including HOI.} (\textbf{a,b}) Zero temperature phase diagram of Pd/Fe/Rh(111) obtained neglecting and including HOI, respectively. The energy of the relaxed spin spiral (SS), the skyrmion lattice (SkX) and ferromagnetic (FM) field-polarized state are shown with respect to the homogeneous spin spiral (black dashed line). The SS, SkX and FM phases are denoted by blue, red and green background color, respectively. \textbf{c-e} Equilibrium spin structures of a SS, a SkX, an isolated skyrmion and an antiskyrmion, respectively. \textbf{g} Radius of isolated skyrmions and antiskyrmions as a function applied magnetic field with respect to the critical field $B_c$ with and without taking HOI into account. Inset of \textbf{g} profile of isolated skyrmions (ISk) and antiskyrmion (IASk) with HOI at $B$= 5.2 T. \textbf{h} Barrier heights of isolated skyrmions and antiskyrmions neglecting and including HOI as a function of magnetic field with respect to $B_c$. Blue and green are used to distinguish two different collapse mechanisms of isolated skyrmions. Above $B-B_{c}$= 1.36 T, isolated skyrmions (including HOI) annihilate by the radial collapse mechanism (solid blue circles) and below $B-B_{c}$= 1.36 T, they annihilate via a chimera structure at the saddle point (solid green circle) and at 1.36 T, they can annihilate by both mechanisms (solid blue-green circle).}
	\label{fig:fig2}
\end{figure*}

\noindent{\textbf{Spin dynamics simulations.}} We use atomistic spin dynamics simulations (see Methods) based on the Hamiltonian of Eq.~(\ref{eq:hamiltonian}) with all parameters obtained from DFT including DMI, MAE and total magnetic moments, as discussed in the previous section, to calculate the zero temperature phase diagram and the properties of isolated skyrmions and antiskyrmions in the three film systems. The results for Pd/Fe/Rh(111) are shown in Fig. \ref{fig:fig2} 
(for the other two film systems see Supplementary Figure 1 and Supplementary Figure 2).

We first discuss the effect of the HOI on the zero temperature phase diagram (Figs.~\ref{fig:fig2}a,b). The FM state (green line) and the homogeneous spin spiral states (solid black line) remain almost unaffected by the higher-order terms. However, the skyrmion lattice (red line) loses a small amount of energy with respect to the homogeneous spin spirals which remains constant throughout the range of magnetic fields. This leads to an expansion of the spin spiral (SS) and FM phases at the expense of the skyrmion lattice (SkX) phase, which is squeezed. Since isolated skyrmions can be stabilized in the FM (field-polarized) phase it is of prime importance. The onset of the FM phase, characterized by the critical field $B_{c}$, has shifted from $2.75$ T to a lower value of $2.25$ T due to the HOI. As expected from the magnetic interaction constants, a similar trend of the phase diagram is obtained for Pd/Fe/Ir(111) (Supplementary Figure 1). However, since the higher-order constants are quite small for Fe/Rh/Re(0001) (cf.~Table \ref{tab:table1}), the phase diagram is basically unchanged
(Supplementary Figure 2).

In our spin dynamics simulations, we have created isolated skyrmions and antiskyrmions in the field-polarized background following the theoretical profile \cite{bocdanov94} and relaxed the spin structures with full set of DFT parameters. The radius of skyrmions and antiskyrmions -- defined as in Ref.~\cite{bocdanov94} -- increases with HOI for Pd/Fe/Rh(111) on average by about 35$\%$ and 30~$\%$, respectively (Fig.~\ref{fig:fig2}g). The skyrmion and antiskyrmion profiles at 5.2 T (see inset of Fig.~\ref{fig:fig2}g) can be fit by the standard skyrmion profile. The antiskyrmion exhibits a steeper profile which reflects that it has a smaller radius than the skyrmion. Similar trends of the skyrmion and antiskyrmion radii are found for Pd/Fe/Ir(111) (Supplementary Figure 1). Due to relatively small HOI, the skyrmion radii remain almost unchanged for low magnetic fields above $B_c$ for Fe/Rh/Re(0001) (Supplementary Figure 2 and Supplementary Note 1).          

To study the stability of metastable isolated skyrmions and antiskyrmions, we employ the geodesic nudged elastic band (GNEB) method \cite{gneb,Bessarab2017prb} which allows to calculate the minimum energy path (MEP) for the collapse of a single skyrmion and antiskyrmion into the FM background (see Methods). The point of maximum energy on this path, known as the saddle point, with respect to the initial state (skyrmion or antiskyrmion) is a measure of the barrier height.
As seen in Fig.~\ref{fig:fig2}h, the HOI increase the energy barrier for skyrmion annihilation in Pd/Fe/Rh(111) by more than a factor of two at small magnetic fields above $B_c$. For antiskyrmions, the barrier height is even increased by a factor of 5. 

The energy barriers of skyrmions vary nonlinearly at small magnetic fields, and, thereafter, reduce almost linearly with increasing magnetic fields \cite{Malottki2017a}. On an average, we notice an increase in barrier height of nearly 100 meV  for skyrmions upon including the HOI. At low fields up to $B-B_c=1.36$~T, there is a transition from the normal radial collapse mechanism \cite{Malottki2017a,som18} without HOI to a chimera collapse mechanism \cite{meyer19,heil19} with HOI. Above $B-B_c=1.36$ T, the skyrmions merge into the FM background through the normal radial collapse without HOI which remains unchanged after including HOI. The barrier heights of antiskyrmions with HOI exhibit a similar variation with field as that of skyrmions. The energy barriers of antiskyrmions without HOI are extremely small ($\sim$ 20 meV) which implies that they are almost unstable. However, after including HOI, the energy barriers become $\sim$100 meV at small fields (up to 1.5~T above $B_c$), which suggests that a metastable antiskyrmion could be realized. The annihilation mechanism of antiskyrmions is via the radial collapse \cite{Malottki2017a}, which is unaffected upon including HOI.    

In Pd/Fe/Ir(111), the HOI increase the energy barriers of isolated skyrmions and antiskyrmions by similar values of
$\sim$100 meV and $\sim$ 90 meV, respectively (Supplementary Figure 1). On the other hand, in Fe/Rh/Re(0001), the stability of isolated skyrmions is reduced on an average by 70 meV upon including the HOI 
(Supplementary Figure 2). This large barrier reduction shows that even small values of the higher-order constants (cf.~table \ref{tab:table1}) can have significant effects.

\begin{figure*}[htbp!]
	\centering
	\includegraphics[scale=0.95,clip]{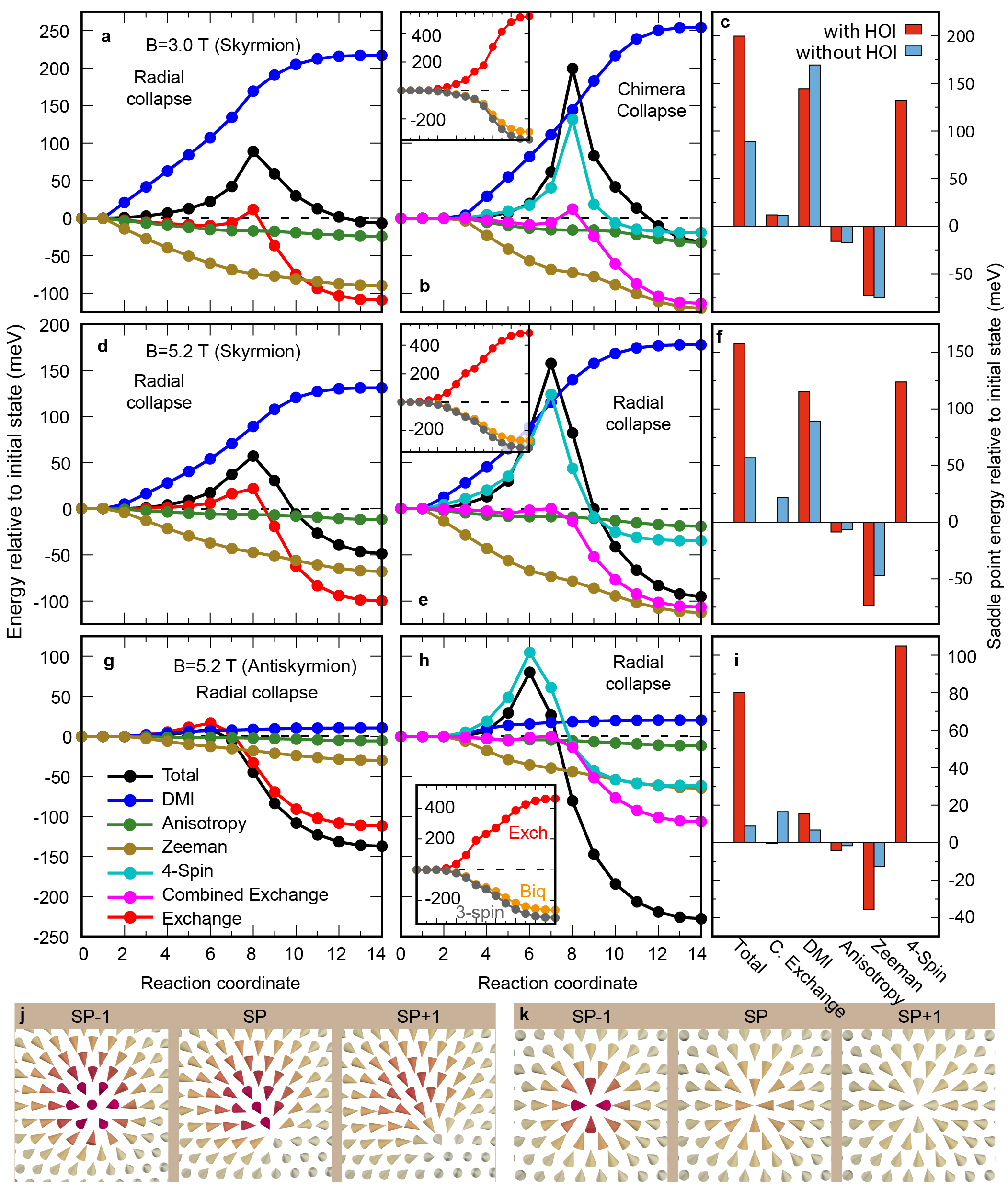}
	\caption{\textbf{Minumum energy paths of skyrmion and antiskyrmion collapse in Pd/Fe/Rh(111)}. Total and individual energy contributions along minimum energy paths neglecting and including HOI. Combined exchange for simulations with HOI denotes the sum of exchange, biquadratic, and three-site four spin interaction. \textbf{a} Radial collapse of an isolated skyrmion at $B=3$~T neglecting HOI. \textbf{b} Chimera collapse of an isolated skyrmion at $B=3$~T including HOI. \textbf{c} Energy decomposition at the saddle point of the path shown in \textbf{a,b} with respect to the initial (skyrmion) state. \textbf{d,e} Radial collapse of an isolated skyrmion at $B=5.2$~T neglecting and including HOI, respectively. \textbf{f} Energy decomposition at the saddle point of the path shown in \textbf{d,e} with respect to the initial (skyrmion) state. \textbf{g,h} Radial collapse of an isolated antiskyrmion at $B=5.2$~T neglecting and including HOI, respectively. \textbf{i} Energy decomposition at the saddle point of the path shown in \textbf{g,h} with respect to the initial (antiskyrmion) state. Inset of \textbf{b,e,h} The exchange, biquadratic and three-site four spin (3-spin) terms. \textbf{j,k} Spin structures before (SP-1), after (SP+1) and at the saddle point (SP) for the chimera collapse in 
  \textbf{b} and the radial collapse in \textbf{e}, respectively. For brevity, the two-site, three-site, and four-site four spin interactions are denoted as Biq, 3-spin and 4-spin, respectively.}
	\label{fig:fig3}
\end{figure*}   

\noindent{{\textbf{Analysis of collapse mechanisms.}}} Now we focus on the question which of the HOI is responsible for the large changes of the energy barriers for skyrmion or antiskyrmion collapse. We consider both collapse mechanisms for skyrmions, i.e., the chimera collapse at low fields and the radial collapse at higher fields and the radial collapse mechanism for antiskyrmions (cf.~Fig.~\ref{fig:fig2}h). In Fig.~\ref{fig:fig3}, the energy decomposition of three 
representative minimum energy paths are displayed at selected magnetic fields for Pd/Fe/Rh(111) with and without taking HOI into account (for the other two film systems similar plots are displayed in Supplementary Figure 3 and Supplementary Figure 4).

The total energy rises along the minimum energy path as one moves from the initial (skyrmion) state to the saddle point and descend thereafter to the final (ferromagnetic) state (Fig.~\ref{fig:fig3}a). In the simulation neglecting the HOI, we find that the energy barrier is dominated by the energy contribution from the DMI which favors the skyrmion state. Due to exchange frustration, there is also a small energy contribution to the barrier from the exchange energy. Naturally, the energy due to the Zeeman term and the magnetocrystalline anisotropy decrease in the ferromagnetic state. 

Upon including the HOI (Fig.~\ref{fig:fig3}b), we find the large increase of the energy barrier as discussed in the previous section. In addition, the annihilation mechanism changes at this magnetic field from the radial collapse without HOI (Fig.~\ref{fig:fig3}a) to the chimera collapse mechanism with HOI (cf.~Figs.~\ref{fig:fig3}j,k) show the spin structures in the vicinity of the saddle point of the two types of annihilation mechanisms). Interestingly, the chimera collapse mechanism has been previously discussed in ultrathin films with very strong exchange frustration \cite{meyer19,heil19,Desplat2019}, which suggests that HOI acts in a similar way. The energy decomposition shows that the DMI contribution is of similar magnitude at the saddle point of the path with and without HOI (Fig.~\ref{fig:fig3}c). However, one cannot compare the exchange interactions before and after the HOI are included in the simulations, since the higher-order terms modify the exchange constants according to Eqs. \ref{eq:impJs}(a-c). Therefore, we add the contributions due to the exchange, the three-site four spin interaction and the biquadratic terms which we denote as combined exchange. The comparison of Fig.~\ref{fig:fig3}a and Figs.~\ref{fig:fig3}b shows that the exchange and the combined exchange behave qualitatively quite similar along the path -- which is also true for the other two collapse mechanisms (Figs.~\ref{fig:fig3}d,e and Figs.~\ref{fig:fig3}g,h). The absolute energy change from exchange to combined exchange at the saddle point is relatively small (Fig.~\ref{fig:fig3}c,f,i). 

The four-site four spin interaction acts in a qualitatively different way compared to all other terms. For all considered paths (Figs.~\ref{fig:fig3}b,e,h), it gains in energy slowly as one approaches the saddle point, in the vicinity of the saddle point it becomes very steep, reaches a maximum at the saddle point and drops quickly thereafter. The
energy contributions at the saddle point (Figs.~\ref{fig:fig3}c,f,i) show that it provides by far the largest difference between the simulations with and without HOI, irrespective of the collapse mechanism or the initial state.

The energy contribution from the three-site four spin and the biquadratic interactions decreases along all collapse processes and the energy drop escalates after the saddle point (see insets of Figs.~\ref{fig:fig3}b,e,h). The difference in energy profile of the DMI and MAE is an associated effect of the HOI caused by the changes in relative spin angles during the collapse process. Figs.~\ref{fig:fig3}(c,f,i) show that the combined exchange can provide a tiny contribution to the energy barrier depending on the collapse mechanism, while the DMI and Zeeman terms assert only a little weight if not compensated by each other. Therefore, the four-site four spin interaction mainly controls change of the barrier height.  

\begin{figure*}[htbp!]
	\centering
	\includegraphics[scale=0.95,clip]{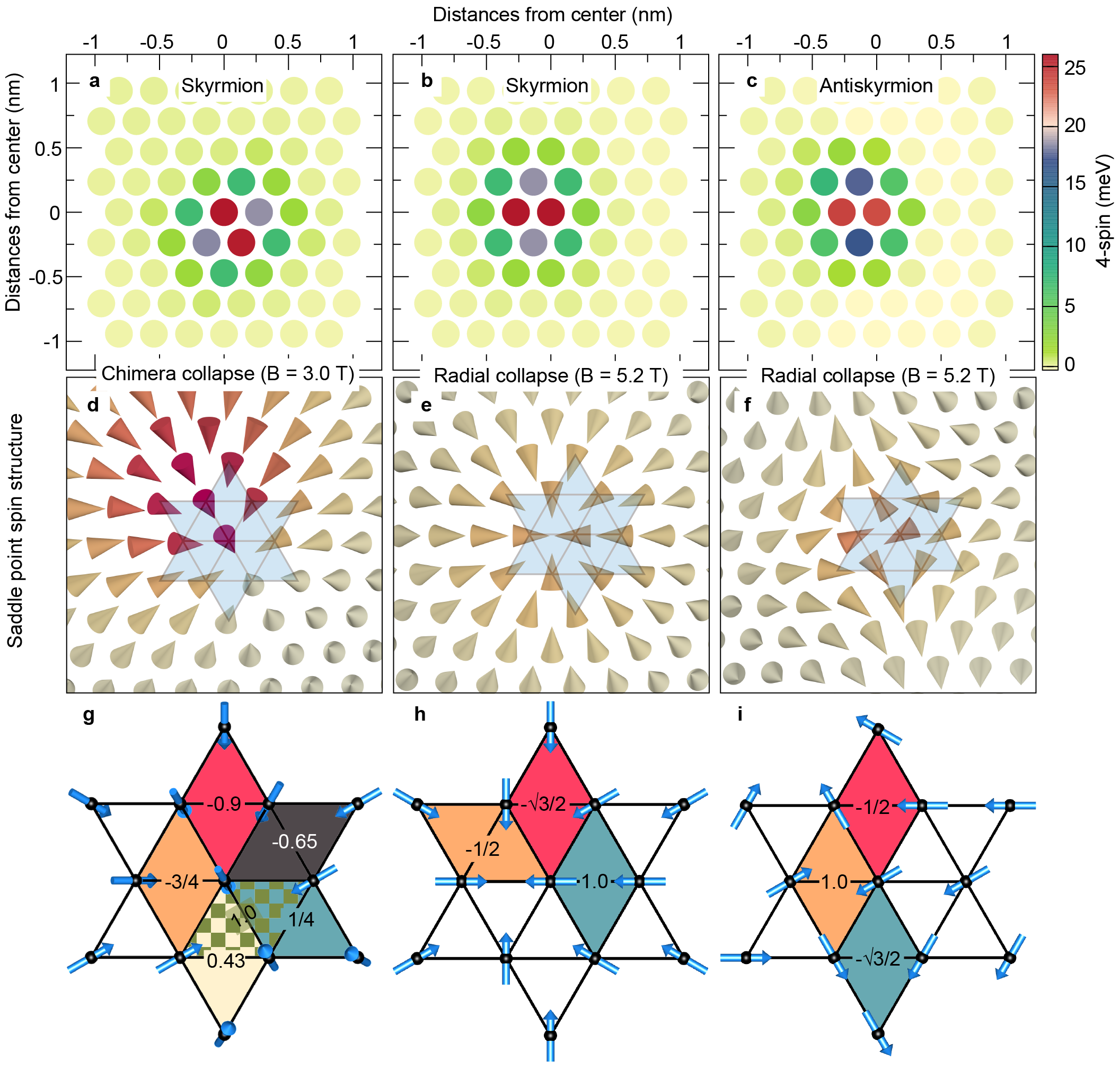}
	\caption{\textbf{Four-site four spin energy at the saddle points of Pd/Fe/Rh(111).} \textbf{a-c} Atomic-site resolved energy contribution of the four-site four spin interaction at the saddle points of the minimum energy paths of Figs.~\ref{fig:fig3}(\textbf{b,e,h}), i.e., for the chimera skyrmion collapse, the radial skyrmion collapse and the radial antiskyrmion collapse. Energies are given with respect to the initial state. \textbf{d-f} Saddle point spin structures of Figs.~\ref{fig:fig3}(\textbf{b,e,h}). \textbf{g-i} Spin structure around the core spin at the origin, i.e., at $(0,0)$, of the saddle point is highlighted in \textbf{d-f} by the gray shaded star. Note that all spins of \textbf{h,i} are in-plane while most of the spins in \textbf{g} are not in-plane. The sign and the value of the four-site four spin interaction is shown for different diamonds.}
	\label{fig:fig4}
\end{figure*} 

The spin structure in the vicinity of the saddle point is shown in Figs.~\ref{fig:fig3}(j,k) for the chimera and the radial collapse of the isolated skyrmion including the effect of HOI. We see that the radial collapse of an isolated skyrmion is very similar to that found by neglecting HOI in Ref. \cite{Malottki2017a}. However, at the saddle point there are four spins pointing towards each other while previously a three-spin structure was reported. The unusual saddle point including HOI is obtained throughout the studied field range and for annihilation of skyrmions in Pd/Fe/Ir(111). However, for the skyrmion collapse in Fe/Rh/Re(0001), a three-spin structure at the saddle point similar to that of Ref.~\cite{Malottki2017a} occurs. The chimera skyrmion collapse and the radial antiskyrmion collapse are similar to that found in simulations neglecting HOI~\cite{meyer19,heil19,Malottki2017a}.

\noindent{{\textbf{Analysis of the four-site four spin interaction}}}. To understand the prominent effect of the four-site four spin interaction on the energy barrier, we present its site-resolved energy at the saddle point with respect to the initial state (skyrmions or antiskyrmions) for the three minimum paths of Figs.~\ref{fig:fig3}(b,e,h) in Figs.~\ref{fig:fig4}(a-c). We notice that a group of only 14 spins around the core provide contributions to the four-site four spin interaction, while the surrounding spins do not add any significant value. This finding is independent of whether we consider the saddle point of the chimera collapse (Fig.~\ref{fig:fig4}a), the radial skyrmion collapse (Fig.~\ref{fig:fig4}b) or the radial collapse of the antiskyrmion (Fig.~\ref{fig:fig4}c). Similar observations are made for the other film systems (Supplementary Figures 5 to Supplementary Figures 8). Therefore, the four-site four spin interaction at the saddle point exhibits a general behavior irrespective of the type of collapse mechanism or the initial spin configuration. 

In order to explain this localized energy gain at the saddle point, we use a simplified model in which we consider only the site with the largest contribution at the origin. To evaluate the four-site four spin interaction, we need to consider at least 6 nearest neighbors and the 6 next-nearest neighbors of the central site [cf.~Fig.~\ref{fig:fig4}(d-f)]. To simplify the discussion, we slightly symmetrize the spin structure. For the radial skyrmion and antiskyrmion collapse, the
12 neighboring spins are nearly all in-plane (Figs.~\ref{fig:fig4}e,f). We neglect any out-of-plane component as shown in Figs.~\ref{fig:fig4}(h,i) to calculate the contributions from the 12 diamonds for the four-site four spin interaction. 

For the saddle point of the radial skyrmion collapse (Fig.~\ref{fig:fig4}h), we find three distinct types of diamonds which contribute to the four-site four spin interaction. We find a pair of diamonds with values $+K_{1}$ and $-K_{1}$, two pairs of diamonds with values $+\frac{1}{2}K_{1}$ and $-\frac{1}{2}K_{1}$, which cancel out mutually. Out of the six remaining diamonds, there are three groups each containing two diamonds with values $-\frac{\sqrt{3}}{2}K_{1}$, $-K_{1}$ and $+\frac{1}{2}K_{1}$, which results in a total energy at the saddle point of 
$E^{\mathrm{ISk}}_{\mathrm{SP}}$= $-2.73K_{1}$. 

For the saddle point of the radial antiskyrmion collapse (Fig. \ref{fig:fig4}i), we identify three types of diamonds with the same magnitude as for the skyrmion saddle point (Fig.~\ref{fig:fig4}h). However, two uncompensated diamonds with $+\frac{1}{2}K_{1}$ and $-\frac{\sqrt{3}}{2}K_{1}$ and two diamonds with $-K_{1}$ each lead to a total energy of $E^{\mathrm{IASk}}_{\mathrm{SP}}$= $-2.37K_{1}$. 

The spin structure at the saddle point of the chimera collapse (Fig.~\ref{fig:fig4}d) is more complex. There are non-negligible out-of-plane components of the spins surrounding the central spin which we take into account in the symmetrization (Fig.~\ref{fig:fig4}g). As a consequence, we find six distinct types of diamonds (Fig.~\ref{fig:fig4}g). Similar to the other two saddle points, there is a mutual cancellation of many terms which leads to a total energy
contribution of the four-site four spin interaction of $E^{\mathrm{chimera}}_{\mathrm{SP}}$= $-2.37K_{1}$.

Note that the values of these three energies taking the exact spin structure at the saddle points are $E^{\mathrm{ISk}}_{\mathrm{SP}}$= $-2.0K_{1}$, $E^{\mathrm{IASk}}_{\mathrm{SP}}$= $-2.2K_{1}$ and $E^{\mathrm{chimera}}_{\mathrm{SP}}$= $-2.14K_{1}$, which are very close to those obtained using the simplified
spin structures.

We obtain similar values for Pd/Fe/Ir(111) at the saddle points corresponding to the skyrmion and antiskyrmion initial states and at the chimera saddle point, the value is only slightly different $E^{\mathrm{chimera}}_{\mathrm{SP}}$= $-$ $2.58K_{1}$ (Supplementary Figures 5 to 7). For Fe/Rh/Re(0001), we find $E^{\mathrm{ISk}}_{\mathrm{SP}}$= $-\sqrt{3}K_{1}$ (Supplementary Figure 8).

The energy contribution per site of the four-site four spin interaction for the ferromagnetic state or any flat spin spiral is $-12 K_1$, which is also relatively close to the energy in the skyrmion state (cf.~Fig.~\ref{fig:fig3}b and Supplementary Figures 3b,e and 4b). Therefore, we obtain an energy difference of $E_{\rm SP}-E_{\rm FM} \approx 10 K_1$ for the two symmetric central sites of the saddle point. The surrounding sites provide smaller contributions, however, they still scale with $K_1$. In total, we find an energy contribution of the four-site four spin interaction to the energy barrier of
roughly $40 K_1$. Due to the linear dependence on $K_1$, it is also clear that the sign of the four-site four spin interaction determines whether there is an energy gain ($K_1>0$) or loss ($K_1<0$) at the saddle point as observed for the two types of systems: Pd/Fe on Rh(111) and on Ir(111) vs. Fe/Rh/Re(0001) (cf.~Table \ref{tab:table1}). 

\begin{figure}[htbp!]
	\centering
	\includegraphics[scale=1.0]{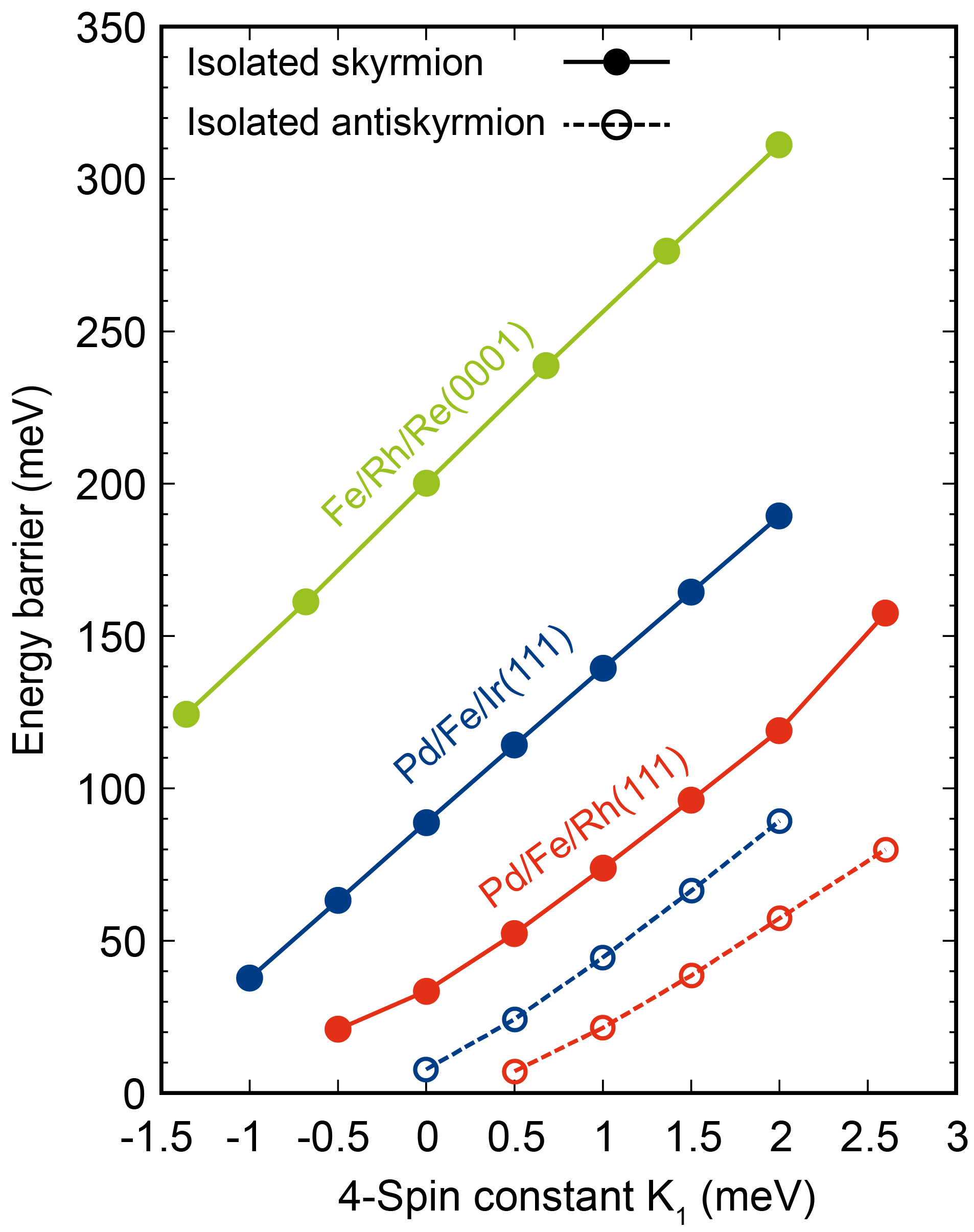}
	\caption{\textbf{Variation of energy barriers with four-site four spin constant}. Energy barriers for skyrmions (filled circles and solid lines) and antisykrmions (open circles and dashed lines) in the three considered ultrathin film systems as a function of the strength and sign of the four-site four spin interaction constant. A constant magnetic field is chosen for each of the systems: $B=5.2$~T for Pd/Fe/Rh(111), $B=6$~T for skyrmions and $B=4$~T for antiskyrmions in Pd/Fe/Ir(111) and $B=3$~T for Fe/Rh/Re(0001).}
	\label{fig:fig5}
\end{figure} 

\noindent{{\textbf{Discussion.}}}
Our simplified model states that the barrier height $E_{\mathrm{SP}}$$-$$E_{\mathrm{ISk/IASk}}$ varies linearly with the magnitude of the four-site four spin constant $K_{1}$. To verify this prediction, we have carried out spin dynamics simulations for three ultrathin film systems at a given magnetic field by changing only the four-site four spin interaction constant while leaving all other interactions the same. Note that the four-site four spin interaction does not affect the energy dispersion of spin spirals which is essential for the equilibrium properties of skyrmions and antiskyrmions. Figure \ref{fig:fig5} shows that -- as expected from our model -- the barrier heights for skyrmions and antiskyrmions exhibit a linear scaling with the four-site four spin constant. Only for Pd/Fe/Rh(111), we find a slight deviations from the linear dependence. The scaling constant $\alpha$, defined as the ratio of the change in energy barrier to the change in four-site four spin constant, is the same for skyrmions and antiskyrmions consistent with our discussion of the energy contributions at the saddle point. We find a value of the scaling constant $\alpha$ of about 40 to 60 depending on the system. 

Figure 3 implies that the HOI can stabilize skyrmions and antiskyrmions by themselves. To test this notion, we have performed spin dynamics simulations by completely switching off the DMI while keeping all other magnetic interactions as before. As shown in Supplementary Figure 9, we find stable skyrmions and antiskyrmions for Pd/Fe bilayers on Rh(111) and Ir(111) with radii reduced to 1.4 and 1.6 nm just above $B_c$, respectively. Significant energy barriers of the order of 80 to 90~meV are obtained due to the four-site four spin interaction (Supplementary Figure 10), which are identical for skyrmions and antiskyrmions as the DMI is zero. This suggests that it is possible to stabilize both types of topological states with opposite rotational sense due to the HOI at inversion symmetric transition-metal interfaces. 

                       
The lifetime $\tau$ of skyrmions is given by the Arrhenius law $\tau$= $\tau_{0}$exp($\Delta E$/$k_{\mathrm{B}}T$), where $\Delta E$ is the energy barrier and $\tau_{0}$ is the prefactor. An enhancement of only the barrier height, $\Delta E$, by 100 meV, as observed for an isolated skyrmion in Pd/Fe/Rh(111), leads to an increase of skyrmions lifetime by orders of magnitude because of the exponential factor. For example, at a temperature of $T=10$ K, at which the spin-polarized scanning tunneling microscopy experiments on such ultrathin films are typically performed \cite{Romming2013,Romming2015}, we find an enhancement by 50 orders of magnitude, at 100 K, it is still a factor of about $10^5$, and even at room temperature, it is a factor of about 50. One can also discuss the effect of the HOI in terms of the temperature dependent phase diagram of Pd/Fe/Rh(111) \cite{som18}. Without the HOI, skyrmions can be stable for an hour up to temperatures of 25~K \cite{som18}, which is increased to a temperature of about 50~K upon including the HOI.
       
\noindent{{\textbf{Conclusion.}}} We have demonstrated that the HOI beyond pair-wise Heisenberg exchange can play a key role for the stability of skyrmions or antiskyrmions at transition-metal interfaces. Depending on the sign of the four-site four spin interaction, the energy barrier preventing the collapse of a metastable topological spin structure can be greatly enhanced or reduced. Even for the small values of HOI typical for $3d$ transition metals, we find large changes of the energy barriers and therefore giant effects on the lifetime, which means that these interactions cannot be neglected. The energy barriers are so enhanced with HOI that it can stabilize isolated skyrmions and antiskyrmions in the absence of Dzyaloshinskii-Moriya interaction. Our study opens up a new avenue to stabilize topological spin structures at transition-metal interfaces.
 
\section*{}
\noindent{\large{\textbf{Methods}}}\par
\noindent{\textbf{First-principles calculations.}} The computational details for calculating the exchange, the DMI, the MAE constants and the magnetic moments are shown in Ref.~\cite{som18} for Pd/Fe/Rh(111), in Ref.~\cite{Malottki2017a} for Pd/Fe/Ir(111) and in Ref.~\cite{mine} for Fe/Rh/Re(0001). Here, we have evaluated the higher-order exchange constants for all three systems. The electronic structure was calculated using a spin-polarized DFT code based on the projected augmented wave (PAW) scheme \cite{paw} as implemented in the Vienna $\textit{ab initio}$ simulation package ($\textsc{vasp}$) \cite{vasp}. It ranks among the best available DFT codes in terms of accuracy and efficiency \cite{dft}. We use the same structural parameters as mentioned in the above references. We have used two atomic overlayers on top of nine substrate layers to mimic the surfaces. To maintain consistency with spin spiral calculations, we have chosen local density approximation (LDA) for the exchange and correlation part of potential \cite{vwn}.  A high energy cut-off of 400 eV was used to precisely calculation the energy of the multi-$Q$ states. The 2DBZ was sampled by a Monkhorst-Pack \cite{mpkmesh} mesh of 22$\times$28$\times$1 $k$-points for the $uudd$ state in the $\overline{\Gamma \mathrm{K}}$ direction, of 14$\times$44$\times$1 $k$-points for the $uudd$ state in the $\overline{\Gamma \mathrm{M}}$ direction and of 15$\times$15$\times$1 $k$-points for the 3$Q$ state at the $\overline{\mathrm{M}}$ point. The total energy convergence criteria were set to 10$^{-6}$~eV for all calculations.

\noindent{\textbf{Fitting function for HOI.} The spin spiral is the exact solution of the classical Heisenberg model for a periodic lattice. The spin spiral, which is characterized by a wave vector $\textbf{q}$ in the 2DBZ and the magnetic moments of an atom at lattice site $\textbf{R}_{i}$, is given by \cite{pkurz},

\begin{gather}
\textbf{M}_{i}=2(\textbf{R}_{\textbf{q}}\mathrm{cos}(\textbf{q} \cdot \textbf{R}_{i})-\textbf{I}_{\textbf{q}}\mathrm{sin}(\textbf{q} \cdot \textbf{R}_{i}))
\end{gather}

where $\textbf{R}_{\textbf{q}}$ and $\textbf{I}_{\textbf{q}}$ are two vectors that span the xy-plane. They obey the following relation,

\begin{gather}
\textbf R^{2}_{\textbf{q}}=\textbf I^{2}_{\textbf{q}}=M^{2}/2, {\textbf R^{2}_{\textbf{q}}} \cdot {\textbf I^{2}_{\textbf{q}}}=0
\end{gather}
 
where $M$ is the magnitude of $\textbf{M}_{i}$ and without loss of generality, we set its norm to unity. The spins of a spin spiral rotate around the z-axis in the xy-plane as one moves from one lattice site to another in the direction of $\textbf{q}$. Using the above two equations, the scalar product of a pair of spins can be written as \cite{pkurz},
\begin{gather} \label{eq3}
\textbf{M}_{i} \cdot \textbf{M}_{j}=\mathrm{cos}(\textbf{q} \cdot \textbf{R})
\end{gather}
In reciprocal space, $\textbf{q}$ is defined as $\textbf{q}$=q$_{x}\textbf{b}_{1}$+$q_{y}\textbf{b}_{2}$, with $\textbf{b}$ being the reciprocal lattice vectors. In our case, $\textbf{b}_{1}$=($2\pi/a$)(1,-1/$\sqrt{3}$) and $\textbf{b}_{2}$=($2\pi/a$)(1,1/$\sqrt{3}$), here $a$ is the in-plane lattice constant. For a spin spiral propagating along $\overline{\Gamma \mathrm{KM}}$, $\textbf{q}$=(2$\pi$/a)q(1,0) with q$\in$[0,1] and along $\overline{\Gamma \mathrm{M}}$ $\textbf{q}$=(2$\pi$/a)q($\sqrt{3}$/2,-1/2) with q$\in$[0,1/$\sqrt{3}$].
 
Using Eq.~(\ref{eq3}), we calculate the energies of the exchange interactions up to third nearest-neighbor, the HOI on a hexagonal lattice along $\overline{\Gamma \mathrm{K}}$ direction as, 
\begin{widetext}
\begin{eqnarray} \label{eq:gkm}
\begin{aligned}
E_{\overline{\Gamma \mathrm{K}}}(J^{\prime}_{1},J^{\prime}_{2},J^{\prime}_{3},Y_{1},B_{1},K_{1})&= -2[\mathrm{cos}(2\pi q)+2\mathrm{cos}(\pi q)][J^{\prime}_{1}+Y_{1}] -2[1+2\mathrm{cos}(3\pi q)][J^{\prime}_{2}+Y_{1}] \\
&-2[\mathrm{cos}(4\pi q)+2\mathrm{cos}(\pi q)][J^{\prime}_{3}+B_{1}/2] -12K_{1}
\end{aligned}
\end{eqnarray}
\end{widetext}
and along $\overline{\Gamma \mathrm{M}}$ direction as,
\begin{widetext}
\begin{eqnarray} \label{gm}
\begin{aligned}
E_{\overline{\Gamma \mathrm{M}}}(J^{\prime}_{1},J^{\prime}_{2},J^{\prime}_{3},Y_{1},B_{1},K_{1})&= -2[1+2\mathrm{cos}(\sqrt{3}\pi q)][J^{\prime}_{1}+Y_{1}] -2[\mathrm{cos}(2\sqrt{3}\pi q)+2\mathrm{cos}(\sqrt{3}\pi q)][J^{\prime}_{2}+Y_{1}] \\
&-2[1+2\mathrm{cos}(2\sqrt{3}\pi q)][J^{\prime}_{3}+B_{1}/2] -12K_{1}
\end{aligned}
\end{eqnarray}
\end{widetext}
  
It is clear from Eqs. (\ref{eq:gkm}) and (\ref{gm}) that we can at most obtain the combined terms $J^{\prime}_{1}+Y_{1}$, $J^{\prime}_{2}+Y_{1}$ and $J^{\prime}_{3}+B_{1}/2$ by fits. Therefore, we evaluate the higher-order exchange constants from the total energy difference between the multi-$Q$ states and single-$Q$ states according to Eqs. 2(a-c) and modify the exchange constants obtained assuming vanishing high-order contributions using Eqs.~\ref{eq:impJs}(a-c) of the main text. The four-site four spin energy is constant, i.e, $-12K_{1}$, for all the spin spiral vector \textbf{q}, which implies that it does not affect the exchange constants.

\noindent{\textbf{Atomistic spin dynamics simulations.} We study the time evolution of atomistic spins as described by Landau-Lifshitz equation, where the dynamics is expressed as a combination of the precession and the damping terms:
	
\begin{gather} \label{eq:llg}
\hslash \frac{d\textbf{m}_{i}}{dt}=\frac{\partial H}{\partial{\textbf{m}}_{i}}\times\textbf{m}_{i}-\alpha\left(\frac{\partial H}{\partial{\textbf{m}}_{i}}\times\textbf{m}_{i}\right)\times\textbf{m}_{i}
\end{gather}	

where $\hslash$ is the reduced Planck constant, $\alpha$ is the damping parameter and the Hamiltonian is $H$ defined in Eq. (1). In the simulation, $\alpha$ has been varied from 0.05 to 0.1, while we have chosen a time step of 0.1 fs and simulated over 2$-$3 millions steps to ensure relaxation of the spin structures. We employed a semi-implicit scheme proposed by Mentink $\textit {et al}$. \cite{mentink2000} to accomplish a time integration of Eq. (\ref{eq:llg}). 
	
\noindent{\textbf{Geodesic nudged elastic band method.}	We calculate the annihilation energy barrier of isolated skyrmions and anitiskyrmions and their collapse mechanism via the geodesic nudged elastic band method (GNEB) \cite{gneb,Bessarab2017prb}. The objective of GNEB is to find a minimum energy path (MEP) connecting initial state (IS), in this case, skyrmion or antiskyrmion, and final state (FS), i.e., FM state, on an energy surface. The GNEB is a chain-of-state method, in which 
a string of images (spin configurations of the system) is used to discretize the MEP. The method selects an initial path connecting IS and FS and systematically brings it to MEP by relaxing the intermediate images. The image relaxation is performed through a force projection scheme, in which the effective field acts perpendicular and spring force acts along the path. The maximum energy on the MEP corresponds to a saddle point (SP) which defines energy barrier separating IS and FS. The energy of the SP is accurately determined using a climbing image scheme on top of GNEB.}

\noindent{\textbf{Data availability.} The data presented in this paper are available from the authors upon reasonable request.

\noindent{\textbf{Code availability.} The atomistic spin dynamics code is available from the authors upon reasonable request.   

\bibliographystyle{naturemag}
\bibliography{reference} 
~\\
\noindent{\textbf{Acknowledgements.} We gratefully acknowledge computing time at the supercomputer of the North-German Supercomputing Alliance (HLRN) and financial support from the Deutsche Forschungsgemeinschaft (DFG, German Research Foundation) via project HE3292/13-1. We thank Pavel F. Bessarab for valuable discussions.

\noindent{\textbf{Author contributions.} S.He. devised the project. S.P. implemented the three-site four spin interaction into the atomistic spin dynamics code and tested the code including HOI.
S.Ha. and S.P. performed the DFT calculations. 
S.Ha., S.P., and S.v.M. performed the spin dynamics and GNEB simulations. S.Ha. and S.P. prepared the figures.   
S.P. and S.He. wrote the manuscript. All authors discussed the data and contributed to preparing the manuscript.   
	
\noindent{\textbf{Additional information.} Supplementary Information accompanies this paper.
			
\noindent{\textbf{Competing interests.} The authors declare no competing financial interests.

\end{document}